\documentclass[11pt]{article}
\usepackage{graphicx}
\usepackage{cite}
\usepackage[table]{xcolor}
\usepackage{booktabs}
\usepackage{multirow}
\usepackage{amsmath,amssymb,mathrsfs}
\usepackage{pstricks}
\usepackage{amsmath,amssymb,subfigure,epsfig,bbm,stmaryrd,MnSymbol,mathrsfs}
\usepackage[margin=3cm]{geometry}

\usepackage{fancyhdr}
\newcommand{\dphi}{\delta\phi}
\newcommand{\dsigma}{\delta\sigma}
\newcommand{\tphi}{\tilde\phi}
\newcommand{\dJ}{\delta J}
\newcommand{\Fb}{\mathbf{F}}
\newcommand{\rb}{\mathbf{r}}
\newcommand{\sigmab}{\boldsymbol{\sigma}}
\newcommand{\de}{\delta e}
\newcommand{\into}{\int_\Omega}
\newcommand{\ds}{d\mathcal{S}}
\newcommand{\dr}{\hspace{.7mm}d\mathbf{r}}

\begin{document}
\lhead{\scriptsize To appear on IEEE Geoscience and Remote Sensing
Letters} \rhead{\scriptsize DOI:
10.1109/LGRS.2011.2164052\normalsize}
\title{\bf{Sensitivity Calculations for the Poisson's Equation via the Adjoint Field
Method}}
\author{Alireza Aghasi,  Eric L. Miller\footnote{Authors are with the Department
of Electrical and Computer Engineering, Tufts University, Medford,
MA 02155, USA. E-mails: \{aaghas01, elmiller\}@ece.tufts.edu. } }

\maketitle

\begin{abstract}
%\boldmath
Adjoint field methods are both elegant and efficient for
calculating sensitivity information required across a wide range
of physics-based inverse problems. Here we provide a unified
approach to the derivation of such methods for problems whose
physics are provided by Poisson's equation. Unlike existing
approaches in the literature, we consider in detail and explicitly
the role of general boundary conditions in the derivation of the
associated adjoint field-based sensitivities. We highlight the
relationship between the adjoint field computations required for
both gradient decent and Gauss-Newton approaches to image
formation. Our derivation is based on standard results from vector
calculus coupled with transparent manipulation of the underlying
partial different equations thereby making the concepts employed
here easily adaptable to other systems of interest.
\end{abstract}



% \fi

\section{Introduction}
\label{sec:introduction}

Geophysical imaging modalities based on the inversion of the
Poisson's equation include electrical resistance tomography (ERT)
\cite{ramirez1993monitoring}, electrical impedance tomography
(EIT) \cite{cheney1999electrical}, and induced polarization (IP)
\cite{oldenburg1994inversion}. Additionally, the same physical
model underlies electrical capacitive tomography (ECT), employed
for nondestructive evaluation
\cite{soleimani2005nonlinear,xie1992electrical} as well as aspects
of the diffuse optical tomography inverse problem arising in brain
and breast imaging \cite{arridge1999optical}. Most all inversion
methods require the calculation of sensitivity information as part
of the imaging algorithms; i.e. the functional derivative of the
cost function or a part of the cost function with respect to the
unknown physical quantity being imaged.

Adjoint field methods represent an analytically elegant as well as
computationally attractive approach for sensitivity calculations
and have been considered for various imaging problems. Among the
main contributions in this area we highlight the work by Somersalo
\emph{et al.} in which the adjoint calculations are analyzed for
the EIT problem using weak forms of the Poisson's equation
\cite{somersalo1992existence}. In the context of electromagnetic
imaging, Dorn \emph{et al.} in \cite{dorn1999nonlinear} have
considered reconstruction of the complex conductivity using
magnetic measurements, for which they use the Maxwell's equations
and derive the required sensitivity information. In an alternative
contribution, shape based reconstruction of the imaging domain is
considered using the differential form of the Helmholtz equation
as the governing modality \cite{dorn2000shape}. Inverse problems
based on the integral forms of the Helmholtz equation and full
wave models taking into account the incident and scattered field
components are another challenging problem for which adjoint
methods play a significant role in simplifying the sensitivity
calculations. For more details about the application of adjoint
technique in these problems, an interested reader is referred to
some of the work of Abubakar \emph{et al.} mainly in the context
of contrast source inversion (CSI)
\cite{abubakar2000nonlinear,abubakar2002contrast,abubakar2004robust}.

While the adjoint field approach has been applied to Poisson
inverse problems, the details of its use depend quite heavily on
the algorithmic method being employed by the inverse solver. More
specifically, the use of a gradient decent-type of approach
\cite{tai2004survey, ben2007projection, dorn2006level} requires a
different adjoint calculation than a Gauss-Newton method for which
a full Jacobian must be determined \cite{lionheart2004eit,
polydorides2002matlab,aghasi2010parametric}. Additionally, it
turns out that given the latter, the former can easily be
obtained.

It is certainly true that the literature contains a number of
similar derivations based on operator-theoretic principals
\cite{chung2005electrical, ben2007projection}, employing weak
formulations
\cite{somersalo1992existence,rylander2010reconstruction,nordebo2010adjoint},
or using physical concepts such as power conservation
\cite{polydorides2002matlab} and the reciprocity theorem
\cite{geselowitz1971application,gunther2006three,abubakar20082}.
While the final, analytical expressions for the sensitivity
information are quite similar (if not the same) to that which is
derived here, the analytical methods used in the various
derivations differ markedly in terms of brevity, clarity, and the
level of mathematical background underlying the analysis.
Moreover, as noted previously, the adjoint computations can be put
to a number of different uses depending on the nature of the
optimization technique being employed to solve the inverse
problem; i.e., gradient decent or Newton-type. The differences in
these uses are not readily apparent in the adjoint-field
derivations for the Poisson problem currently available in the
literature.

The primary contribution of this paper is the presentation of a
clear and detailed derivation based on easily accessible, vector
calculus identities (following one of Norton's methods in
\cite{norton1999iterative} where the Helmholtz equation was
considered) of adjoint based sensitivity calculations for Poisson
based inverse problems. We highlight the mathematical relationship
between the gradient descent and Newton-type adjoint forms and
consider problems with general type of boundary conditions,
including Neumann, Dirichlet or mixed boundary conditions. While
we certainly acknowledge that some of the results presented in the
paper are well known, we feel that the pedagogical approach we
have taken, our comprehensive and detailed treatment of boundary
conditions, and the explicit treatment of the calculations
required for gradient decent and Newton-type uses of the approach
represent a contribution to the literature. Moreover, by employing
a minimum of mathematical abstraction and a step-by-step
derivation of all results, we hope that the approach we take here
can provide the interested reader with the tools needed to apply
the ideas to problems governed by other physical models.

\section{General Problem Formulation}
For an imaging domain $\Omega$ with surface $\Gamma$, the forward
model of interest motivated by several geophysical applications is
\begin{align}
\nabla \cdot (\sigma \nabla \phi)&=s &
\mbox{in}\;\Omega,\label{eq:1}
\\\alpha\sigma\frac{\partial \phi}{\partial
\mathbf{n}}+\beta\phi&=0 &\mbox{on}\;\Gamma.\label{eq:2}
\end{align}
In the context of electrical tomography as a basic modality,
$\phi$ is the electrical potential, $s$ current source and
$\sigma$ the conductivity. All quantities are assumed to be
functions of three spatial coordinates, $\rb = [x\; y\; z]^T$.
Regarding the boundary condition (\ref{eq:2}), the notation
$\partial \phi /\partial \mathbf{n}$ denotes the normal component
of the gradient of $\phi$ on $\Gamma$, and $\alpha(\rb)$ and
$\beta(\rb)$ are functions defined on the surface $\Gamma$ (i.e.,
$\rb\in\Gamma$) which are not simultaneously zero. To maintain
generality and consider problems with different types of boundary
conditions on different regions of $\Gamma$, we do not make any
continuity assumptions about $\alpha$ and $\beta$. Also,
generalization of the results to nonhomogeneous boundary condition
will be accomplished later in this letter.

\begin{figure}[t]\label{fig1}
\hspace{-1.6cm}\epsfig{file=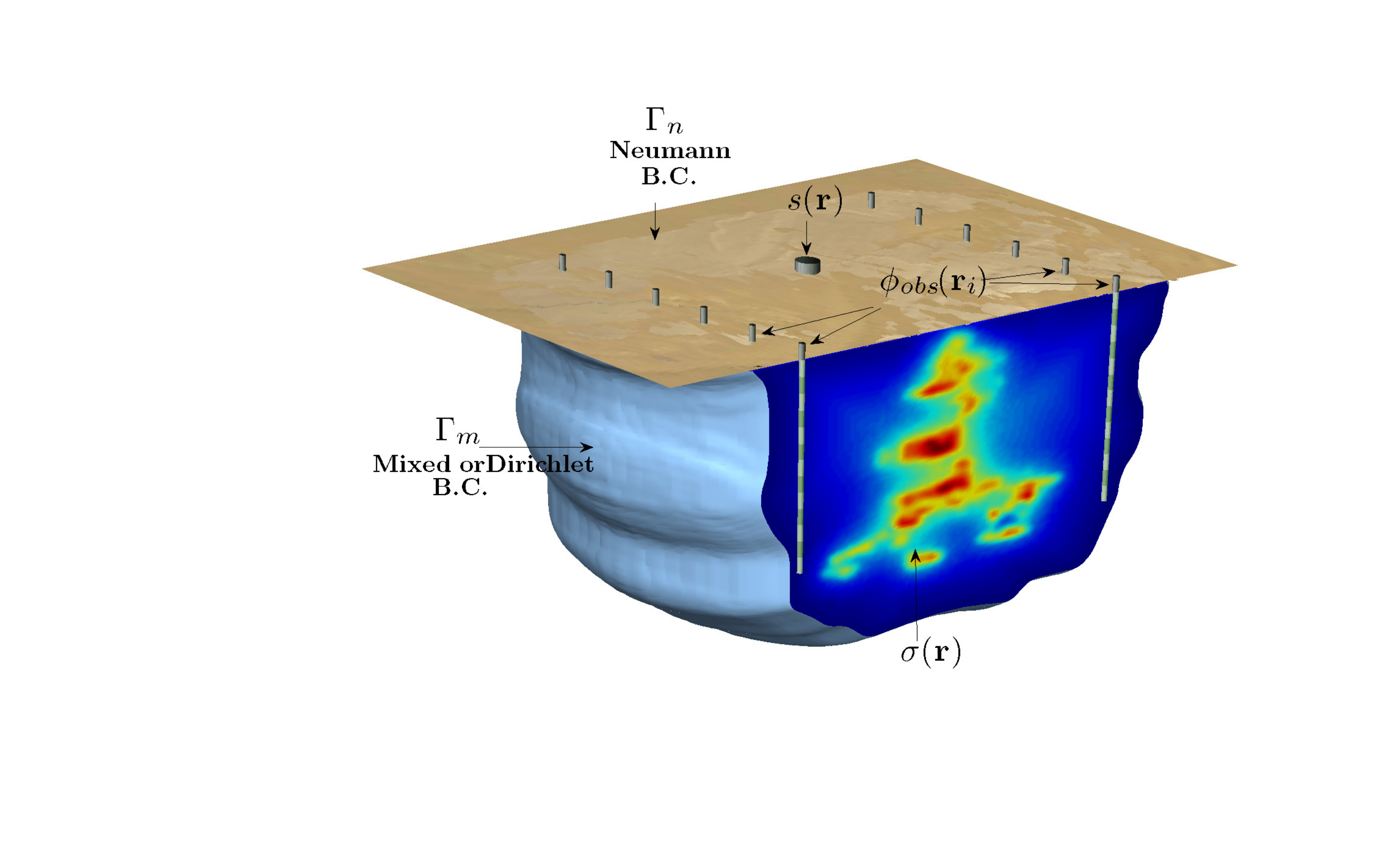,width=1\linewidth,clip=a}\\\vspace{-2cm}
\caption{An ERT application: reconstruction of the subsurface
conductivity distribution based on the potential measurements at
inserted boreholes. A Neumann boundary condition (B.C.) considered
for the top surface and a mixed or Dirichlet type B.C. considered
for the remaining boundaries}
\end{figure}

A common geophysical problem associated with the model in
(\ref{eq:1})--(\ref{eq:2}) is shown in Fig. 1. In this problem
$\Gamma$ consist of $\Gamma_n$, the interface between the earth
and air where we impose a zero current condition ($\beta=0$) and
$\Gamma_m$, a surface where the values $\alpha$ and $\beta$ are
chosen to model an infinite half-space
{\cite{pollock2008temporal}. As an approximation to an infinite
half-space, the boundary condition on $\Gamma_m$ may be replaced
with a zero potential condition ($\alpha=0$) when $\Gamma_m$ is
far from the sources of current \cite{tripp1984two}.
Reconstruction of $\sigma$ based on the measurements of $\phi$ at
some points in the domain is the goal of this tomography problem.

For simplicity here, we consider only real-valued conductivity
(that is the ERT problems) although the approach can easily be
adapted for use in estimation of complex valued conductivity as is
encountered in EIT as well as conductivity/chargability as are
desired in IP experiments. To keep the notation simple, we also
consider the case where data are collected from a single source of
current. More generally, in a tomographic imaging problem one
would illuminate with many sources $s_p$ for $p=1,2,\cdots,P.$

Central to the derivation that follows is the impact of a small
change in conductivity to the system (\ref{eq:1})--(\ref{eq:2}).
Consider a perturbation to the conductivity, $\sigma\rightarrow
\sigma+\dsigma$ resulting in $\phi\rightarrow \phi+\dphi$. Using
these in (\ref{eq:1})--(\ref{eq:2}) gives
\begin{align}
\nabla \cdot \big((\sigma+\dsigma) \nabla (\phi+\dphi)\big)&=s & \mbox{on}\;\Omega,\label{eq:3}\\
\alpha(\sigma+\dsigma)\frac{\partial }{\partial
\mathbf{n}}(\phi+\dphi)+\beta(\phi+\dphi)&=0
&\mbox{on}\;\Gamma.\label{eq:4}
\end{align}
Now, expanding (\ref{eq:3}) as
\begin{equation*}
\nabla\cdot(\sigma\nabla\phi)+\nabla\cdot(\dsigma\nabla\phi)+\nabla\cdot(\sigma\nabla\dphi)+\nabla\cdot(\dsigma\nabla\dphi)=s,
\end{equation*}
keeping terms of first order and using (\ref{eq:1}) in
(\ref{eq:3}) and (\ref{eq:2}) in (\ref{eq:4}) yields
\begin{align}
\nabla \cdot (\dsigma \nabla \phi)+\nabla \cdot (\sigma \nabla \dphi)&=0, &  \mbox{on}\;\Omega,\label{eq:5}\\
\alpha(\sigma\frac{\partial}{\partial
\mathbf{n}}\dphi+\dsigma\frac{\partial \phi}{\partial
\mathbf{n}})+\beta\dphi&=0 &\mbox{on}\;\Gamma.\label{eq:6}
\end{align}
The goal of the inverse problem is to estimate $\sigma$ from
observations of the potential collected at a finite set of points
in space $\rb_i$, $i = 1,2,\cdots,N$. The estimate is obtained by
minimizing the least squares cost function
\begin{align}\nonumber
J(\sigma)&=\frac{1}{2}\sum_{i=1}^N\big(\phi(\rb_i)-\phi_{obs}(\rb_i)\big)^2\\
&=\frac{1}{2}\;\mathbf{E}(\sigma)^T\mathbf{E}(\sigma),\label{eq:7}
\end{align}
where $\phi_{obs}(\rb_i)$ are the observations of potential at
location $\rb_i$ and $\phi(\rb_i)$ is implicitly a function of
$\sigma$ through (\ref{eq:1})--(\ref{eq:2}). The length $N$ column
vector $\mathbf{E}(\sigma)$ contains the $N$ residuals,
$e_i(\sigma)=\phi(\rb_i)-\phi_{obs}(\rb_i)$ for $i =
1,2,\cdots,N$. A few remarks are in order regarding (\ref{eq:7}):
\begin{itemize}
\item The use of a least squares formulation is not terribly
restrictive. For example, a one-norm type of cost function could
be accommodated by populating $\mathbf{E}$ with a suitably
smoothed version of $|\phi(\rb_i)-\phi_{obs}(\rb_i)|^{1/2}$.
Though somewhat tedious, the results in this letter could then be
generalized. \item Generalization of (\ref{eq:7}) to account for
weighting of the residuals is also not difficult. To avoid the
resulting notational burdens, we choose to not include these
details. \item In many cases, the cost function includes both a
data misfit term as in (\ref{eq:7}) as well as additional
regularization. We refer the reader to
\cite{tikhonov1977solutions} for details concerning the treatment
of Tikhonov-type regularization schemes. The adjoint field
calculations of interest here pertain only to the data term.
\end{itemize}

As discussed in Section \ref{sec:introduction}, two type of
sensitivities are of interest depending on the nature of the
inversion algorithm. Methods based on gradient information such as
the steepest decent, require the functional derivative of the cost
function with respect to $\sigma$. More formally, a linear
integral operator is sought which maps small perturbations of the
conductivity, $\dsigma$, to corresponding changes in $J$. In the
case of Newton type methods and more specifically the Gauss-Newton
approach, one requires $N$ linear operators relating $\dsigma$ to
perturbations in the individual residuals, $\delta e_i$. We start
with the latter. Subsequently we indicate the changes to the
derivation required to obtain the former and also how the adjoint
calculations used in the Newton type approaches can be assembled
to obtain the gradient-based sensitivity information.

\section{Adjoint Field Calculations}
\label{sec:adjo-field-calc}

To start, we consider the variation in $e_i$ due to a small change
in $\sigma$. We have
\begin{equation*}
  \de_i(\sigma) =  \dphi(\rb)|_{\rb=\rb_i},
\end{equation*}
which can be written as a volume integral
\begin{equation}
  \label{eq:8}
  \de_i(\sigma) =  \into \delta(\rb-\rb_i)\dphi(\rb)\dr,
\end{equation}
where $\delta(.)$ denotes the Dirac delta function. The objective
here is to find the linear integral operator relating $\de_i$ to
$\dsigma$. Toward this end we define the $i$-th adjoint source
$\tilde{s}_i$ as
\begin{displaymath}
  \tilde{s}_i(\rb) = \delta(\rb-\rb_i).
\end{displaymath}
It is now useful to define the potential, $\tphi_i$, as the
solution to the adjoint system
\begin{align}
\nabla\cdot(\sigma\nabla\tphi_i) & = \tilde{s}_i & \mbox{on}\;\Omega,\label{eq:9}\\
\alpha\sigma \dfrac{\partial \tphi_i}{\partial
\mathbf{n}}+\beta\tphi_i & =
  0 & \text{ on } \Gamma.\label{eq:10}
\end{align}
Physically, $\tphi_i$ is the potential field arising from the
solution to the adjoint Poisson's equation (which in this case is
the same as the original since Poisson's equation is self-adjoint)
due to a point source of amplitude unity located at the $i$-th
receiver.  From (\ref{eq:8}) and (\ref{eq:9}) we conclude that the
perturbation to the residuals can be written in terms of the
adjoint field as
\begin{equation}
  \label{eq:11}
  \de_i(\sigma) =  \into \nabla\cdot(\sigma\nabla\tphi_i)
  \dphi(\rb)\dr.
\end{equation}
The goal is to find a collection of kernel functions, $k_i(\rb)$
such that (\ref{eq:11}) can be expressed as
\begin{equation}
  \label{eq:12}
  \de_i(\sigma) =  \into k_i(\rb) \dsigma(\rb)\dr.
\end{equation}
The resulting kernels are known as the Frech\'et derivative of
$e_i$ with respect to $\sigma$ \cite{dieudonn1969foundations}.

To find these kernels, we make extensive use of the following
identity derived from Green's theorem \cite{marsden2003vector} for
vector function $\Fb$ and scalar function $g$
\begin{equation}
  \label{eq:13}
  \into \Fb\cdot\nabla g \dr + \into g(\nabla\cdot\Fb)\dr  =
  \int_\Gamma
  g\Fb\cdot d\mathcal{S},
\end{equation}
where $\int \mathbf{V}\cdot \ds=\int (\mathbf{V}\cdot
\mathbf{n})\;dS$ denotes the surface integral over $\Gamma$ of the
normal component of the vector field $\mathbf{V}$.

We begin by taking $g = \dphi$ and $\Fb = \sigma\nabla\tphi_i$ in
(\ref{eq:11}) to obtain
\begin{equation}
  \label{eq:14}
  \de_i(\sigma) =
    -\into \sigma(\nabla\tphi_i)\cdot (\nabla\dphi)\dr
  + \int_\Gamma \sigma\nabla\tphi_i\dphi\cdot \ds.
\end{equation}
Next using $g = \tphi_i$ and $\Fb = \sigma\nabla\dphi$ in the
first term on the right hand side of (\ref{eq:14}), we have
\begin{align}
  \nonumber
    \de_i(\sigma) &=
     \into \tphi_i \nabla\cdot(\sigma\nabla\dphi) \dr\\\label{eq:15}
    &- \int_\Gamma \sigma(\nabla\dphi)\tphi_i\cdot \ds
    + \int_\Gamma \sigma\nabla\tphi_i\dphi\cdot \ds.
\end{align}
From (\ref{eq:5}), $\nabla\cdot\dsigma\nabla\phi = -
\nabla\cdot\sigma\nabla\dphi$ which we use in the first term on
the right hand side of (\ref{eq:15}) to arrive at
\begin{align}
  \nonumber
  \de_i(\sigma) =&
     -\into \tphi_i \nabla\cdot(\dsigma\nabla\phi) \dr\\\label{eq:16}
    &- \int_\Gamma \sigma(\nabla\dphi)\tphi_i\cdot \ds
    + \int_\Gamma \sigma(\nabla\tphi_i)\dphi\cdot \ds.
\end{align}
Appealing once more to (\ref{eq:13}) with $g = \tphi_i$ and $\Fb =
\dsigma\nabla\phi$ in the first term of (\ref{eq:16}) gives
\begin{align}
  \label{eq:17}
  \de_i(\sigma) &=
     \into \dsigma (\nabla \tphi_i)\cdot(\nabla\phi)
      \dr\\\nonumber
    &- \int_\Gamma \Big(\sigma(\nabla\dphi)\tphi_i+\dsigma(\nabla \phi) \tphi_i-\sigma(\nabla\tphi_i)\dphi\Big)\cdot
    \ds.
\end{align}
We now rewrite the surface integral term on the right hand side of
(\ref{eq:17}) as the integration over the normal component
\begin{equation}\label{eq:18}
\mathscr{I}=\int_\Gamma \Big( \sigma\frac{\partial}{\partial
\mathbf{n}}\dphi\tphi_i+\dsigma\frac{\partial \phi}{\partial
\mathbf{n}}\tphi_i -\sigma \dfrac{\partial \tphi_i}{\partial
\mathbf{n}}\dphi\Big) dS,
\end{equation}
and show that it is zero. For this purpose we multiply both sides
of (\ref{eq:10}) by $\dphi$ to arrive at
\begin{equation}
  \label{eq:19}
  \beta\dphi\tphi_i+\alpha\sigma \dfrac{\partial \tphi_i}{\partial
\mathbf{n}}\dphi =
  0.
\end{equation}
Using (\ref{eq:6}) to replace the term $\beta\dphi$ in
(\ref{eq:19}) results in
\begin{equation}
  \label{eq:20}
  -\alpha \Big( \sigma\frac{\partial}{\partial
\mathbf{n}}\dphi\tphi_i+\dsigma\frac{\partial \phi}{\partial
\mathbf{n}}\tphi_i -\sigma \dfrac{\partial \tphi_i}{\partial
\mathbf{n}}\dphi\Big)=0, \qquad \text{on}\;\Gamma.
\end{equation}
The expression within the brackets in (\ref{eq:20}) is the same as
the integrand in (\ref{eq:18}). Considering this term, if
$\alpha\neq 0$ for $\Gamma_\alpha \subset \Gamma$, then clearly
the inside bracket expression becomes zero on $\Gamma_\alpha$. For
the remaining surface $\Gamma\setminus\Gamma_\alpha$ that
$\alpha=0$, we certainly have $\beta\neq 0$ since $\alpha$ and
$\beta$ may not be simultaneously zero and using this fact in
(\ref{eq:6}) and (\ref{eq:10}) would result in $\dphi=0$ and
$\tphi_i=0$ which again make the inside bracket term zero.
Therefore for all $\rb\in\Gamma$ the inside bracket term is zero
and hence the surface integral in (\ref{eq:18}) vanishes,
resulting in
\begin{equation}
  \label{eq:21}
  \de_i(\sigma) =   \into \dsigma (\nabla \tphi_i)\cdot(\nabla\phi)
  \dr.
\end{equation}
This result expresses the perturbation to $e_i$ as a linear
operator acting on $\dsigma$ as desired, and accordingly
identifying the kernel function in (\ref{eq:16}) as $k_i(\rb) =
\big(\nabla \tphi_i(\rb)\big)\cdot\big(\nabla\phi(\rb)\big)$.

In case of a system with nonhomogeneous boundary condition
\begin{align}
\nabla \cdot (\sigma \nabla \phi)&=s &
\mbox{in}\;\Omega,\label{eq:22}
\\\alpha\sigma\frac{\partial \phi}{\partial
\mathbf{n}}+\beta\phi&=\xi &\mbox{on}\;\Gamma,\label{eq:23}
\end{align}
perturbing $\sigma\rightarrow \sigma+\dsigma$ and $\phi\rightarrow
\phi+\dphi$ would result in the same equations as
(\ref{eq:5})--(\ref{eq:6}) and $\xi$ will be cancelled. Based on
this result, by solving the same forward problem as
(\ref{eq:9})--(\ref{eq:10}) and finding $\tphi_i$, identical forms
of sensitivity as (\ref{eq:21}) will be obtained for the case of
nonhomogeneous boundary condition. However, we should note that
although the sensitivity results are valid, by definition of an
adjoint system, (\ref{eq:9})--(\ref{eq:10}) is not the adjoint of
the nonhomogeneous system (\ref{eq:22})--(\ref{eq:23}). More
details in this regard are provided in the Appendix.

In cases where gradient decent-type optimization methods are used
for imaging, one requires the functional derivative of the cost
function, $J(\sigma)$ in (\ref{eq:7}) with respect to the
conductivity. The required variation now is
\begin{equation*}
  \dJ(\sigma) = \sum_{i=1}^N
  \big(\phi(\rb_i)-\phi_{obs}(\rb_i)\big)\dphi(\rb_i),
\end{equation*}
which similarly may be rewritten in an integral form as
\begin{align*}
\dJ(\sigma) &= \sum_{i=1}^N
\big(\phi(\rb_i)-\phi_{obs}(\rb_i)\big)\into \delta(\rb-\rb_i)\dphi(\rb)\dr\\
&=\into\!\! \Big(\! \sum_{i=1}^N\!
\big(\phi(\rb_i)-\phi_{obs}(\rb_i)\big)
\delta(\rb-\rb_i)\!\Big)\dphi(\rb)\dr.
\end{align*}
There are two ways to determine the resulting Frech\'{e}t
derivative. On the one hand, if we define the composite adjoint
source
\begin{equation}
  \label{eq:24}
  \tilde{s}(\rb) =\sum_{i=1}^N \big(\phi(\rb_i)-\phi_{obs}(\rb_i)\big)
\delta(\rb-\rb_i)
\end{equation}
the derivation provided above follows unaltered and we conclude
\begin{equation}
  \label{eq:25}
  \dJ(\sigma) = \into \dsigma (\nabla \tphi)\cdot(\nabla\phi) \dr,
\end{equation}
where now $\tphi$ is a single adjoint field computed according to
(\ref{eq:9})--(\ref{eq:10}) with $\tilde{s}$ replacing
$\tilde{s}_i$.  Thus, for gradient decent methods, one requires
only two solutions of Poisson's equation: one for the source and
one for the composite adjoint source as opposed to $1+N$ solves
needed for a Gauss-Newton scheme.  Alternatively, since $\tilde{s}
= \sum_i \big(\phi(\rb_i)-\phi_{obs}(\rb_i)\big) \tilde{s}_i$, by
the linearity of Poisson's equation and (\ref{eq:24})
\begin{align*}
  \dJ(\sigma) &= \sum_{i=1}^N \big(\phi(\rb_i)-\phi_{obs}(\rb_i)\big) \de_i(\sigma)\\ \nonumber &=
  \sum_{i=1}^N \big(\phi(\rb_i)-\phi_{obs}(\rb_i)\big) \into \dsigma (\nabla \tphi_i)\cdot(\nabla\phi)
  \dr,
\end{align*}
which is the desired link between the two uses for the adjoint
field in sensitivity calculations.

As a concrete example, consider a 3D imaging problem where
$\sigma(\rb)$ is discretized to form a vector $\sigmab=[\sigma_1,
\cdots, \sigma_M]^T$. Here the $j$-th element corresponds to a
voxel $\Omega_j$ in the domain, over which the conductivity is
assumed to be uniformly $\sigma_j$. To iteratively minimize the
least squares problem (\ref{eq:7}) using a gradient descent
approach, the $k$-th iteration to find $\sigmab^{(k+1)}$ is
performed by moving along the negative direction of $\nabla J$ as
\begin{align*}
\sigmab^{(k+1)}-\sigmab^{(k)}&=-\gamma^{(k)}\nabla
J^{(k)}\\&=-\gamma^{(k)}(\nabla\mathbf{E}^{(k)})^T\mathbf{E}^{(k)}
\end{align*}
where $\gamma^{(k)}>0$ is a step size
\cite{bertsekas1999nonlinear}. On the other hand, for Newton type
methods, the updating is performed through solving the following
system for $\sigmab^{(k+1)}$
\begin{equation*}
\mathbf{B}^{(k)}(\sigmab^{(k+1)}-\sigmab^{(k)})=-(\nabla\mathbf{E}^{(k)})^T\mathbf{E}^{(k)},
\end{equation*}
where $\mathbf{B}$ is a positive definite approximation to the
Hessian of $J$, such as $(\nabla \mathbf{E})^T\nabla\mathbf{E}$ in
a Guass-Newton approach or $(\nabla
\mathbf{E})^T\nabla\mathbf{E}+\gamma \mathbf{I}$ in a
Levenberg-Marquardt case \cite{bertsekas1999nonlinear}. Clearly,
in using a gradient descent approach only $\nabla J$ (basically
the product of $\nabla \mathbf{E}^T\mathbf{E}$) is required at
every iteration while in a Gauss-Newton approach $\nabla
\mathbf{E}$ itself is required. The $(i,j)$-th element of the $N
\times M$ matrix $\nabla \mathbf{E}$ is $\partial e_i/\partial
\sigma_j$. By taking $\dsigma$ in (\ref{eq:21}) to be
$\dsigma_j\chi_{\Omega_j}(\rb)$ with $\chi_A$ the indicator
function over a set $A$, we have
\begin{equation*}
  \frac{\partial e_i}{\partial\sigma_j} =
    \int_{\Omega_j} (\nabla \tphi_i)\cdot(\nabla\phi) \dr.
\end{equation*}
Thus, the cost to calculate $\nabla \mathbf{E}$ is solving one
forward problem to compute $\phi$, $N$ adjoint problems to
determine $\tphi_i$, and, for each $i$ the evaluation of $M$
integrals to determine each row of the matrix. However, in a
gradient descent approach where only the vector $\nabla J$ of
length $N$ is required each element may be calculated through
(\ref{eq:25}) as
\begin{equation*}
  \frac{\partial J}{\partial\sigma_j} =
    \int_{\Omega_j} (\nabla \tphi)\cdot(\nabla\phi) \dr,
\end{equation*}
which requires solving one forward problem to compute $\phi$, one
adjoint problem to obtain $\tphi$ and $N$ integrations.

\section{Conclusion}
\label{sec:conclusions} We presented explicit and generalizable
methods of calculating the sensitivity in inversion of Poisson's
type problems. For problems that minimize the misfit between the
data and the model for the purpose of inversion, gradient descent
methods or Newton type methods such as the Gauss-Newton and the
Levenberg-Marquardt may be used. From an implementation point of
view, gradient descent methods are easy to implement, but
iteratively slow and have variable scaling issues
\cite{bertsekas1999nonlinear}. On the other hand Newton type
methods have a faster convergence rate and are robust to variable
scalings but can be computationally expensive. Based on the nature
of the problem and the available computing resources either
methods may be desirable in solving an inverse problem. It is
highlighted that to maintain efficiency, two different adjoint
problems need to be solved to obtain the sensitivity information
in each inversion scheme. This paper beside providing a step by
step derivation of the sensitivities for the Poisson's equation
with general type of boundary condition, clarifies the distinction
and the relationship between the two forms of inversion for
various types of applications.

\section{Appendix} By definition, for the systems
(\ref{eq:22})--(\ref{eq:23}) and (\ref{eq:9})--(\ref{eq:10}) to be
adjoints we must have
\begin{equation}\label{eq:26}
\into \nabla\cdot(\sigma\nabla\tphi_i)
  \phi\dr =\into \nabla\cdot(\sigma\nabla\phi)
  \tphi_i\dr.
\end{equation}
Applying the identity (\ref{eq:13}) twice to $\into
\nabla\cdot(\sigma\nabla\tphi_i)
  \phi\dr$, first with
$g=\phi$ and $F=\sigma\nabla \tphi_i$, and next with $g=\tphi_i$
and $F=\sigma\nabla \phi$ results in
\begin{align}\nonumber
  \into \nabla\cdot(\sigma\nabla\tphi_i)
  \phi\dr -&\into \nabla\cdot(\sigma\nabla\phi)
  \tphi_i\dr\\= &\int_\Gamma \Big(\sigma \nabla \tphi_i\phi - \sigma \nabla \phi\tphi_i\Big)\cdot
  \ds\label{eq:27}
\end{align}
We next split the surface integral on the right hand side of
(\ref{eq:27}) over regions $\Gamma_\alpha$ where $\alpha\neq 0$
and $\Gamma \setminus \Gamma_\alpha$ where $\alpha=0$. On
$\Gamma_\alpha$ using (\ref{eq:10}) and (\ref{eq:23}) we have
\begin{align}
\sigma \frac{\partial\tphi_i}{\partial\mathbf{n}}\phi-\sigma
\frac{\partial\phi}{\partial\mathbf{n}}\tphi_i
=-\frac{\xi\tphi_i}{\alpha},\label{eq:28}
\end{align}
and over $\Gamma\setminus\Gamma_\alpha$ that $\alpha=0$ and
$\beta\neq0$, using (\ref{eq:10}) and (\ref{eq:23}) yields
$\tphi_i=0$ and $\phi=\xi/\beta$ and therefore
\begin{align}
\sigma \frac{\partial\tphi_i}{\partial\mathbf{n}}\phi-\sigma
\frac{\partial\phi}{\partial\mathbf{n}}\tphi_i
=\frac{\sigma\xi}{\beta} \frac{\partial\tphi_i}{\partial
\mathbf{n}}.\label{eq:29}
\end{align}
Based on (\ref{eq:28}) and (\ref{eq:29}) we rewrite (\ref{eq:27})
as
\begin{align}\label{eq:30}
  \into \nabla\cdot(\sigma\nabla\tphi_i)
  \phi\dr -&\into \nabla\cdot(\sigma\nabla\phi)
  \tphi_i\dr\\\nonumber= &-\int_{\Gamma\setminus \Gamma_\alpha}
  \frac{\xi\tphi_i}{\alpha}dS+\int_{\Gamma_\alpha}
  \frac{\sigma\xi}{\beta} \frac{\partial\tphi_i}{\partial
\mathbf{n}}dS.
\end{align}
Comparing (\ref{eq:30}) to (\ref{eq:26}) shows that for the
systems (\ref{eq:22})--(\ref{eq:23}) and
(\ref{eq:9})--(\ref{eq:10}) to be adjoints the expression on the
right hand side of (\ref{eq:30}) needs to be zero and this is not
generally the case. Clearly (\ref{eq:26}) holds for the
homogeneous case ($\xi=0$) and the two systems are adjoints as
mentioned in the text.
\section*{Acknowledgment}
This work is supported by the National Science Foundation under
the grant EAR 0838313.



\end{document}